\documentclass[lettersize,journal]{IEEEtran}
\usepackage{amsmath,amsfonts}
\usepackage{algorithmic}
\usepackage{algorithm}
\usepackage{array}
\usepackage[caption=false,font=normalsize,labelfont=sf,textfont=sf]{subfig}
\usepackage{textcomp}
\usepackage{stfloats}
\usepackage{url}
\usepackage[colorlinks=true, linkcolor=blue, citecolor=blue, urlcolor=blue]{hyperref}
\usepackage{verbatim}
\usepackage{graphicx}
\usepackage{cite}
\usepackage{bm} 
\usepackage{tabularx}
\usepackage{tcolorbox}
\usepackage[utf8]{inputenc}
\title{\fontsize{22}{24}\selectfont Joint Beamforming and Integer User Association using a GNN with Gumbel-Softmax Reparameterizations}

\author{Qing Lyu,~\IEEEmembership{Student Member,~IEEE,} and
Mai Vu,~\IEEEmembership{Senior Member,~IEEE,}

\thanks{Copyright (c) 2025 IEEE. Personal use of this material is permitted. However, permission to use this material for any other purposes must be obtained from the IEEE by sending a request to pubs-permissions@ieee.org. 

Qing Lyu and Mai Vu are with the Department of Electrical and Computer Engineering,
Tufts University, Medford, MA 02155 USA (e-mail: (qing.lyu@tufts.edu, mai.vu@tufts.edu). This work was supported in part by the National Science Foundation under CNS grant 2340284.}
}

\makeatletter
\renewcommand\subsection{\@startsection{subsection}{2}{\z@}%
  {0.1\baselineskip} 
  {0.2\baselineskip} 
  {\normalfont\normalsize\itshape}}
\makeatother

\begin{document}

\maketitle

\begin{abstract}
Machine learning (ML) models can effectively optimize a multi-cell wireless network by designing the beamforming vectors and association decisions. Existing ML designs, however, often needs to approximate the integer association variables with a probability distribution output. We propose a novel graph neural network (GNN) structure that jointly optimize beamforming vectors and user association while guaranteeing association output as integers. The integer association constraints are satisfied using the
Gumbel-Softmax (GS) reparameterization, without increasing computational complexity. Simulation results demonstrate that our proposed GS-based GNN consistently achieves integer association
decisions and yields a higher sum-rate, especially when 
generalized to larger networks, compared to all other fractional
association methods.
\end{abstract}

\vspace{-1em}
\section{Introduction}
\IEEEPARstart{M}{illimeter} wave communication  for modern 5G/6G wireless networks provides a promising solution to the spectrum shortage and ever-increasing capacity demand. Beamforming is imperative to overcome strong pathloss in such systems and deliver high network sum-rate. In addition, user association plays an important role in resource allocation for cellular networks, as base stations (BSs) placement in an mmWave network becomes dense to compensate for the strong pathloss at higher frequency bands. An optimal user association helps maintain high network performance. 

Both beamforming and user association optimization are non-linear and non-convex, which has high computational workload and often requires relaxation to convert the problem into a convex form for easier solutions. For beamforming optimization, well-known algorithms such as the weighted minimum mean squared error (WMMSE) \cite{WMMSE} converges within a few iterations but is still computationally expensive because of the matrix inversion in each iteration. User association problems are integer programming, which is known to be NP-hard. Low-complexity distributed algorithms such as in \cite{UAFractional} can converge to a near-optimal solution, but produce fractional user association instead of integer values. 

To facilitate real-time implementation, GNNs have shown great potential because of their simple feed-forward computation and scalability to unseen problem sizes, as a GNN can be employed in networks much larger than the one used for training without the need to retrain \cite{2018relational}. The existing ML-based works, however, cannot output integer association decision directly. For example, the integer association constraint is relaxed to continuous values to solve for user association and beam selection in \cite{deng2023gnn_aided}. Similarly, a relaxed integer constraint is used to jointly optimize beam selection and link activation for device-to-device  networks in \cite{he2022gblinks}. However, relaxing the integer constraint can introduce bias and hence is likely to produce sub-optimal association decisions when converted back to integer values. Furthermore, these works consider a beam codebook which may not be best suitable for dynamic communication networks.  

To allow integer outputs  while still retaining gradient for backpropagation in a ML model, we can apply certain reparameterization techniques. The Straight-Through Gumbel-Softmax (STGS) is a reparameterization method designed to handle categorical variables, allowing them to be trained using gradient-based optimization methods \cite{fan2022soft, jang2017categorical}. STGS provides a differentiable approximation of sampling from a categorical distribution, which is essential when working with discrete variables in neural networks. A Gumbel–Softmax (GS) approach was employed for discrete element selection in a reconfigurable intelligent surface (RIS) using a convolutional neural network (CNN) \cite{revirewer_TCOM}. However, their problem setting of single cell RIS element selection and the employed CNN structure make the GS method not directly applicable to our multicell user association problem using GNNs.

In this work, we proposed an efficient GNN structure with STGS reparameterization for joint, unsupervised optimization of continuous beamforming and integer user association. The proposed GNN can always produce strictly integer association decisions, without converting them from a probability output as in $\operatorname*{softmax}$ or using relaxation. Our proposed GNN offers strong scalability and generalization with network size, and achieves a higher network sum-rate than all existing integer relaxation methods.

\vspace{-1em}
\section{System Model}
\vspace{-0.3em}
\subsection{System and Signal Models}
We consider a downlink multicell system where $M$ BSs, each equipped $N$ antennas, serve $K$ single-antenna user equipments (UEs). Denote the user association matrix as $\bm{A} \in \mathbb{R}^{K \times M}$, where the association vector between BS $m$ and $K$ UEs is represented by column as $\bm{a}_m$ with integer elements $a_{k,m} \in \{0,1\}$, if $a_{k,m}=1$, UE $k$ is associated with BS $m$. Assume unique association where each UE can only connect to one BS, each row of $\bm{A}$ sums to 1. 

Denote the beamforming matrix at BS $m$ as $\mathbf{V}_{m} \in \mathbb{C}^{N \times K}$ where each column $\mathbf{v}_{m,k}$ is the beamforming vector for UE $k$. The transmitted signal from BS $m$ can be written as 
\begin{align}
    \mathbf{x}_m = \bm{a}_m^T\mathbf{V}_m \mathbf{s} = \sum_{k=1}^Ka_{k,m}\mathbf{v}_{m,k}s_k
\end{align}
where $\mathbf{s} \in \mathbb{C}^{K \times 1}$ is the transmitted symbols for $K$ UEs with $E[\mathbf{s}\mathbf{s}^*]=\mathbf{I}$.
The received signal at UE $k$ is 
\begin{align*} 
   y_k = & \sum_{m=1}^M a_{k,m}\mathbf{h}_{m,k}^T \mathbf{v}_{m,k} s_k + \sum_{m=1}^M\sum_{l=1, l\neq k}^K a_{l,m}\mathbf{h}_{m,k}^T \mathbf{v}_{m,l} s_l +n_k
\end{align*}
where $\mathbf{h}_{m,k} \in \mathbb{C}^{N \times 1}$ represents the channel vector from BS $m$ to UE $k$, and $n_k$ is the noise at UE $k$ following $\mathcal{CN}(0, \sigma_k^2)$. The signal-to-interference-noise ratio (SINR) at UE $k$ is 
\begin{align} \label{SNIR}
    \gamma_k = \frac{|\sum_{m=1}^Ma_{k,m}\mathbf{h}_{m,k}^T \mathbf{v}_{m,k}|^2}{\sum^K_{l=1,l\neq k}|\sum^M_{m=1}a_{l,m}\mathbf{h}^H_{m,k}\mathbf{v}_{m,l}|^2 + \sigma_k^2}
\end{align}
The achievable sum-rate is then
\begin{align} \label{sum-rate}
    R = \sum^K_{k=1} \log_2(1+ \gamma_k) \quad \text{(bps/Hz)}
\end{align}

\subsection{Problem Formulation}
The problem of maximizing the sum-rate with integer association and power constraints can be formulated as 

\begin{subequations}
\renewcommand{\theequation}{\theparentequation\alph{equation}}
\setcounter{equation}{-1}
\begin{align} 
    \max_{\bm{A}, \mathbf{V}} \quad & \sum_{k=1}^{K} \log_2(1+ \gamma_k) \label{objective}  \\
    \text{s.t.} \quad & \sum_m a_{k,m} = 1, \quad 
    a_{k,m} \in \{0,1\} \label{constraint_integer}  \\ 
    & \sum_{k=1}^K||a_{k,m}\mathbf{v}_{m,k}||^2=P_m, m=1, ..., M \label{beampower_constraint}
\end{align}
\end{subequations}
Constraint (\ref{constraint_integer}) ensures that each UE can only associate with one BS, and captures the power constraint per BS where each BS can have a unique power budget. 

\vspace{-0.7em}
\section{Edge-update GNN architecture}

In communication networks, the number of UEs or BSs can change over time. Designing adaptable ML models is crucial, but traditional fully connected deep neural networks with fixed input and output sizes will require resizing and retraining, which is inefficient and impractical. GNNs can leverage network structures and reuse per-node and per-edge functions, allowing them to handle varying network sizes. They support combinatorial generalization and can operate on graphs of different sizes than what they are trained on, as shown in \cite{2018relational}. As such, we propose a GNN structure to solve for the user association and beamforming problem in (\ref{objective}). 

\subsection{Graph structure for the wireless network}
\begin{figure}[!t]
\centering
\includegraphics[height=1.43in]{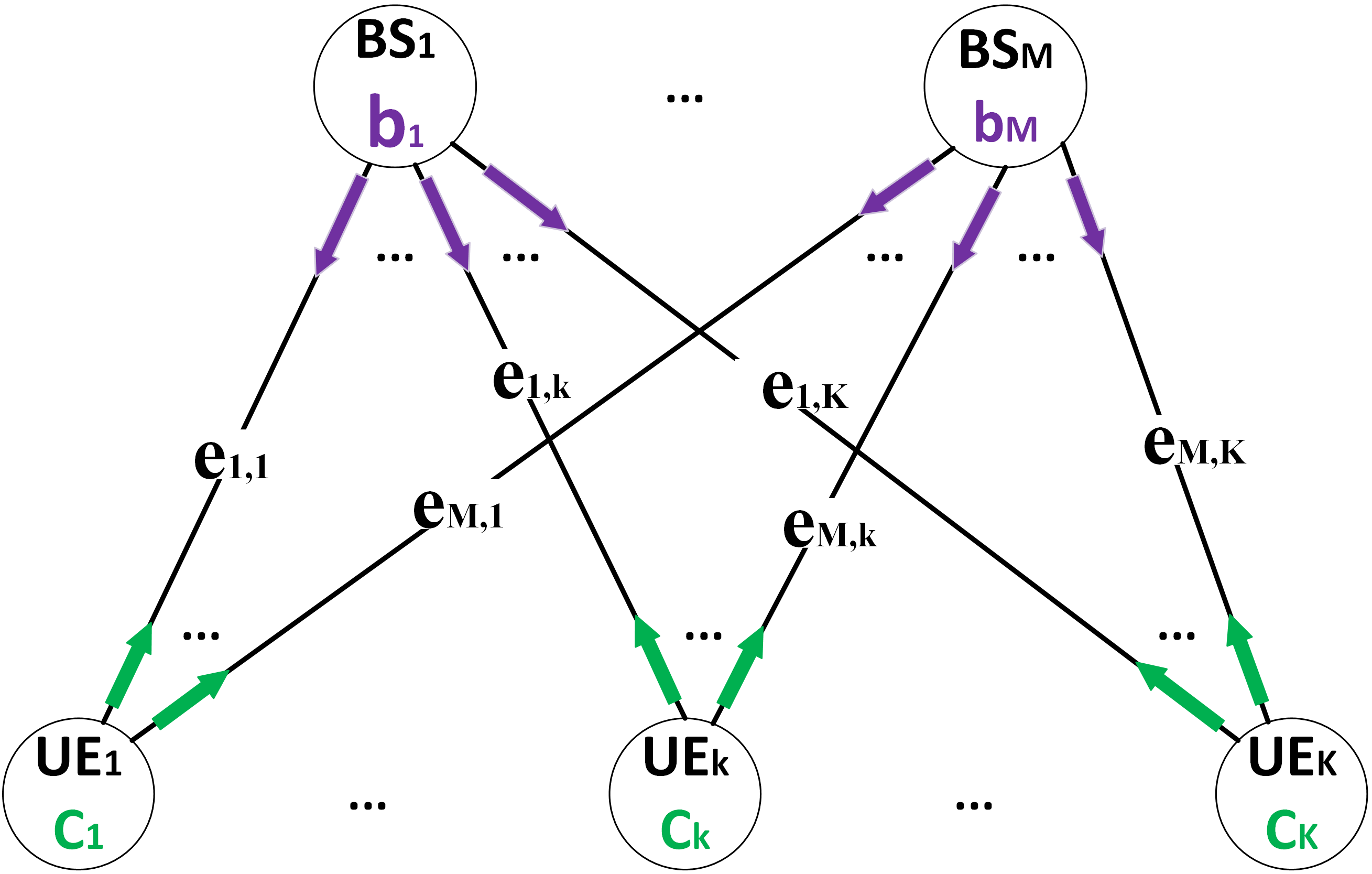}
\caption{The bipartite graph for proposed GNN structure.}
\label{GNN_graph}
\end{figure}

The wireless network can be modeled as a bipartite graph, where the BSs and UEs are two types of nodes, with no edges between nodes of the same type. Such a graph can be represented as $\mathcal{G}=\{\mathcal{M}, \mathcal{K}, \mathcal{E}\}$, where $\mathcal{M}$ is the set of BS nodes, $\mathcal{K}$ is the set of UE nodes, and $\mathcal{E}$ is the set of edges with $\mathcal{E} \stackrel{\triangle}{=} \{(m,k)\}_{m \in \mathcal{M}, k \in \mathcal{K}}$ as shown in Fig. \ref{GNN_graph}. Each edge models the communication link between a BS and a UE and has a representation $\mathbf{e}_{m,k}$. Each BS $m$ has a representation $\mathbf{b}_m$ and each UE $k$ has a representation $\mathbf{c}_k$. All the representations are updated during the GNN training process to facilitate learning to maximize the network sum-rate. Here the UE node representations are converted to user association factors and the edge representations are converted to beamforming vectors. 

Each BS has a power budget $P_m$ and each UE has noise power $\sigma_k^2$ as the input features. The channel $\mathbf{h}_{m,k}$ is used as the input feature for each edge. Since the channel coefficients are complex, the real and imaginary parts are cascaded to create the real-valued input vectors to the GNN as 
\begin{align} \label{edge feature}
    \mathbf{H}_{(m,k,:)} = [\Re\{\mathbf{h}_{m,k}\}^T, \Im\{\mathbf{h}_{m,k}\}^T]^T
\end{align}
where $\mathbf{H} \in \mathbb{R}^{M \times K \times 2N}$.

\subsection{Edge-update GNN structure}
Instead of the typical learning in GNN at the nodes only \cite{RIS, deng2023gnn_aided}, we consider learning at both the nodes and the edges, since edges of a graph generated based on the wireless network structure can represent useful information. Next, we present our proposed edge-update GNN consisting of three types of layers. The preprocessing layer takes the powers and real-value channel coefficients as inputs to produce the initial node/edge representations. The update layers then update these representations during the learning process. Finally, the postprocessing layer converts the UE node representations to association factors and edge representations to beamforming vectors. The method for generating the integer user association factors is discussed in Section \ref{section: STGS}. 

\subsubsection{Preprocessing Layer}
Denote the input features for $M$ BSs as $\mathbf{p} \in \mathbb{R}^{M}$ and that for $K$ UEs as $\mathbf{q} \in \mathbb{R}^{K}$. The preprocessing layer transforms the inputs $(\mathbf{p}, \mathbf{q}, \mathbf{H})$ into the initial node and edge representations $\mathbf{B}^{(0)} \in \mathbb{R}^{M \times d_{BS}}, \mathbf{C}^{(0)} \in \mathbb{R}^{K \times d_{UE}}, \mathbf{E}^{(0)}\in \mathbb{R}^{M \times K \times d_{E}}$ using three multi-layer perceptrons (MLPs), where $d_{BS}, d_{UE}, d_{E}$ are the representation sizes on BS-nodes, UE-nodes and edges, respectively. 

\subsubsection{Update Layer}
To update the node and edge representations, the $l$-th updating layer takes $(\mathbf{B}^{(l-1)}, \mathbf{C}^{(l-1)}, \mathbf{E}^{(l-1)})$ as the inputs, and output the updated node and edge representations $(\mathbf{B}^{(l)}, \mathbf{C}^{(l)}, \mathbf{E}^{(l)})$ as follows. 

\paragraph{Node Update} When updating the representation of BS $m$, the inputs include the previous layer's representation of itself, and the aggregation of all neighboring UEs and edges:
\begin{align}
    \mathbf{b}_m^{(l)} = f_2^{(l)}\Big(\mathbf{b}_m^{(l-1)}, \phi_{\text{BS}}^{(l)}(f_1^{(l)}(\mathbf{c}_k^{(l-1)},
    \mathbf{e}_{m,k}^{(l-1)}))_{k \in \mathcal{N}_m^{\text{UE}}})\Big)\label{BS_representation}
\end{align}
where $\mathbf{b}_{m}^{(l-1)} \overset{\triangle}{=} \mathbf{B}_{(m,:)}^{(l-1)}$,  $\mathbf{c}_{k}^{(l-1)} \overset{\triangle}{=} \mathbf{C}_{(k,:)}^{(l-1)}$, $\mathbf{e}_{m,k}^{(l-1)} \overset{\triangle}{=} \mathbf{E}_{(m,k,:)}^{(l-1)}$ and $\mathcal{N}_m^{\text{UE}}$ denotes the set of neighboring UEs of BS $m$. $f_1$ and $f_2$ are two MLPs, and $\phi_{\text{BS}}^{(l)}$ is an aggregation function (e.g., mean or max aggregation). 

Similarly, the representation update for UE $k$ in the $l$-th updating layer is as follows.
\begin{align}
    \mathbf{c}_k^{(l)} = f_4^{(l)}\Big(\mathbf{c}_k^{(l-1)}, \phi_{\text{UE}}^{(l)}(f_3^{(l)}(\mathbf{b}_m^{(l-1)},
    \mathbf{e}_{m,k}^{(l-1)}))_{m \in \mathcal{N}_k^{\text{BS}}}\Big) 
    \label{UE_representation}
\end{align}
where $f_3$ and $f_4$ are two new MLPs and $\mathcal{N}_k^{\text{BS}}$ is the set of neighboring BSs of UE $k$. 

\paragraph{Edge Update} The edge update mechanism is:
\begin{align}
    \mathbf{e}_{m,k}^{(l)} = f_6^{(l)}\Big(\mathbf{e}_{m,k}^{(l-1)}, \phi_\text{E}^{(l)}\{ f_5^{(l)}(\mathbf{b}_m^{(l-1)}, \mathbf{c}_k^{(l-1)})\}\Big)
    \label{edge_representation}
\end{align}
where $f_5$ and $f_6$ are two MLPs. This edge-update mechanism aggregates information from both nodes connected to either end node of the current edge. Specifically, to update the representation of the edge between BS $m$ and UE $k$, we cascade the representations of BS $m$ and UE $k$ from the previous layer $(\mathbf{b}_m^{(l-1)}, \mathbf{c}_k^{(l-1)})$, and feed each pair into $f_5$. Then we cascade aggregation of this MLP output with the same edge representation from last layer and feed it into $f_6$ to produce the edge representation for the current layer. This edge update mechanism is efficient while capturing all relevant information.

\subsubsection{Postprocessing Layer} \label{postprocessing}
After updating the representations for $L$ layers in GNN, the resulting UE representation matrix $\mathbf{B}^{(L)}$ is converted to integer-based user association matrix $\bm{A}$, as elaborated in Section \ref{integer-based rho}. The edge representation $\mathbf{E}^{(L)}$ is then projected into the feasible region of beamforming power constraint, as in Eq. (\ref{beampower_constraint}). Before projection, we process the edge representation $\mathbf{E}^{(L)}$ using an MLP to reshape it into a size appropriate for beamforming, resulting in $\mathbf{\widetilde{V}} \in \mathbb{R}^{M \times K \times 2N}$. 
The beamforming projection is then performed to satisfy the BS power constraint (\ref{beampower_constraint}) as
\begin{align} \label{AV}
   &a_{k,m}\mathbf{v}_{m,k} = \\ \nonumber
   &\frac{\sqrt{P_m} a_{k,m}[\mathbf{\widetilde{v}}_{m,k}^T(1:N)+j\mathbf{\widetilde{v}}_{m,k}^T(N+1:2N)]}{\sqrt{\sum_{k=1}^K a_{k,m}||\mathbf{\widetilde{v}}_{m,k}(1:N)+j\mathbf{\widetilde{v}}_{m,k}(N+1:2N)||^2}}
\end{align}%
\noindent Note that this projection step is important as it not only uses integer association output $\bm{A}$, but also ensures beamforming optimization only for the associated users as specified by $a_{k,m}$ and allows implicit power allocation among the resulted beamforming vectors.

\subsection{GNN Training}
We employ a mini-batch training approach, where gradient updates are performed after accumulating and averaging gradients from all samples in each mini-batch. Each sample consists of a set of BS transmit powers, noise powers, and cascaded real-value channel coefficients. The objective function in Eq. (\ref{objective}) is regarded as the loss function for training:
\begin{align} \label{loss}
    L{(\mathbf{\Omega})}=-\mathbb{E}_{\text{batch}}[R]
\end{align}
where $\mathbf{\Omega}$ is the set of all MLPs parameters in the GNN.

\vspace{-0.4em}
\section{Gumbel-softmax Reparameterization for integer association} \label{section: STGS}
Integer-based user association is challenging as the optimization problem is NP hard. Existing work using traditional optimization methods often only solve for fractional (continuous) association \cite{UAFractional, deng2023gnn_aided}, which is sub-optimal due to the relaxation of integer constraint. ML methods also cannot output integer values directly, as these values have zero gradients during backpropagation. $\operatorname*{Softmax}$ is the most common technique to convert continuous ML outputs to categorical outputs. However, it only produces a probability vector, rather than always producing integer outputs. 

GS reparameterization is a technique that uses a temperature parameter to make the output approach integer values. Combining GS with other techniques such as Straight-Through estimation can guarantee integer outputs in the feedforward while retaining gradients in the backpropagation. Next, we describe these methods and their applications in our proposed GNN structure.

\subsection{Gumbel-Softmax Reparameterization}
The GS reparameterization addresses the difficulty of differentiating discrete variables in neural networks. Given a categorical distribution vector $\mathbf{d}$, its direct optimization is challenging due to the discrete nature. The goal is to enable backpropagation by making the process differentiable. GS achieves this by transforming the discrete variable into a deterministic, differentiable function of the distribution's parameters combined with a fixed Gumbel-distributed noise \cite{fan2022soft}. This enables gradient-based optimization for ML models with discrete choices.

Starting with the Gumbel-Max method which aims to generates a categorical vector $\mathbf{d} \in \mathbb{R}^M$ from a continuous, unnormalized input vector $\bm{\beta} \in \mathbb{R}^M$ parameterized by $\bm{\theta}$ in a neural network. The vector $\mathbf{d}$ is generated as 
\begin{align} \label{Gumbel_max}
    \mathbf{d} = \operatorname*{argmax} (\text{log} (\bm{\beta}) + \mathbf{g}))
\end{align}
where $\mathbf{g}$ is an i.i.d $M$-dimensional Gumbel(0,1) random vector whose elements can be generated as $-\text{log}(-\text{log}(u))$, with $u \sim \text{Uniform}(0,1)$. However, Eq. (\ref{Gumbel_max}) is not differentiable due to $\operatorname*{argmax}$, hence the GS method is used to remedy this. The GS distribution is defined as a continuous distribution over the simplex that can approximate samples from a categorical distribution \cite{fan2022soft, jang2017categorical}.
Specifically, the method involves using the $\operatorname*{softmax}$ function as a continuous and differentiable approximation to $\operatorname*{argmax}$, and generating an $M$-dimensional sample vector $\mathbf{d}_\text{GS} \in \mathbb{R}^M$ as
\begin{align} \label{d_GS}
    \mathbf{d}_\text{GS} = \operatorname*{softmax}((\text{log}(\bm{\beta}) + \mathbf{g})/\tau)
\end{align}
where vectors $\bm{\beta}$ and $\mathbf{g}$ are as described in Eq. (\ref{Gumbel_max}) and $\tau \in (0, \infty)$ is a temperature factor. As $\tau \rightarrow 0$, samples from Eq. (\ref{d_GS}) approach one-hot coding which is integer.

\subsection{Straight-Through Estimator}
Despite bringing back the differentiability, GS in
Eq. (\ref{d_GS}) still produces fractional output and does not guarantee an integer/categorical output. This can be solved by adding the Straight-Through estimator, where the gradient is not calculated through the standard chain rule, but instead through a modified chain rule in which the derivative of the identity function serves as a proxy of the original derivative of the integer output function \cite{fan2022soft}. Combining these operations, the Straight-Through Gumbel-Softmax (STGS) can always have categorical outputs but still maintain differentiability.

Specifically, let $\mathbf{d}_\text{1hot} = \text{OneHot}(\mathbf{d}_\text{GS})$ be the one-hot coded version of $\mathbf{d}$. The STGS introduces a constant shift of $\mathbf{d}_\text{GS}$
\begin{align} \label{d_STGS}
    \mathbf{d}_\text{STGS} = \mathbf{d}_\text{1hot} - [\mathbf{d}_\text{GS}]_{\text{const}} + \mathbf{d}_\text{GS}
\end{align}
where $[\mathbf{d}_\text{GS}]_{\text{const}}$ is a newly introduced variable which has the same value as $\mathbf{d}_\text{GS}$ but does not depend on the neural network parameters $\bm{\theta}$. In the forward direction, the value of $[\mathbf{d}_\text{GS}]_{\text{const}}$ will cancel out the value of $\mathbf{d}_\text{GS}$, leaving only the integer-valued $\mathbf{d}_\text{1hot}$ as output. In the backward direction, however, only $\mathbf{d}_\text{GS}$ contributes to gradient computation, as neither $[\mathbf{d}_\text{GS}]_{\text{const}}$ nor $\mathbf{d}_\text{1hot}$ depends on $\bm{\theta}$ Therefore, $\mathbf{d}_\text{STGS}$ consists of one-hot coded integers in the forward propagation and is differentiable in the backward propagation \cite{fan2022soft}.

\subsection{Reparameterization for Integer Association Factors} \label{integer-based rho}
We now discuss the application of the above GS and STGS methods to ensure integer association output in our GNN structure. For both GS and STGS, although the gradients are based on $\mathbf{d}_\text{GS}$, the discrete output $\mathbf{d}_\text{1hot}$ produced by STGS in the feedforward process may have an effect of forcing the GNN model to reduce its exploration during training. Contrarily, the fractional output of GS may encourage the GNN model to explore and learn more robustly during training, which could be beneficial for generalization \cite{2018relational}. As such, we explore the use of both methods: GS and STGS in the postprocessing stage of our GNN as follows.

For UE $k$, its last update layers's representation $\mathbf{c}_k^{(L)}$ is converted to the unnormalized parameter $\bm{\beta}_k \in \mathbb{R}^M$ as 
\begin{align}
    \bm{\beta}_k = |\text{f}_7 (\mathbf{c}_k^{(L)})|
\end{align}
This vector $\bm{\beta}_k$ is then used to generate the vector $\mathbf{d}_\text{GS}$ as in Eq. (\ref{d_GS}). 
Each element in $\mathbf{d}_\text{GS} \in \mathbb{R}^M$ corresponds to the probability of UE $k$ being associated with BS $m$. The elements in $\mathbf{d}_\text{GS}$ are not necessarily integers. We convert $\mathbf{d}_\text{GS}$ to integer association output as follows.

\subsubsection{Gumbel-Softmax}
In this method, we use Eq. (\ref{d_GS}) in the GNN to maintain differentiability and fractional association for training, but use its one-hot coded version for the sum-rate evaluation. Specifically, the integer association vector for UE $k$ $\hat{\bm{a}}_{k}^T=\mathbf{d}_\text{1hot}$ is used for computing the sum-rate as in Eq. (\ref{sum-rate}), while the fractional association $\bm{a}_k^T = \mathbf{d}_\text{GS}$ is used in the loss function for the GNN training process as in Eq. (\ref{loss}).

\subsubsection{Straight-Through Gumbel-Softmax}
In this method, we use Eq. (\ref{d_STGS}) in the GNN to obtain the integer association factors $\bm{a}_k^T = \mathbf{d}_\text{STGS}$, which directly satisfy the integer constraint (\ref{constraint_integer}). The same association vector $\bm{a}_k^T$ is used in computing both the network sum-rate in Eq. (\ref{sum-rate}) and the loss function for GNN training in Eq. (\ref{loss}).

\subsection{Comparative Analysis of STGS and GS}
Although both the GNN models with STGS and GS reparameterizations are optimized with the same loss function in Eq. (\ref{loss}), they differ fundamentally in how the intermediate association vector is represented, and this difference lead to distinct learning dynamics and generalization behavior. 
With GS alone, the feedforward process produces continuous-value associations. Because they are differentiable, any small change in the association vector produces a proportional change in the loss, yielding a smooth, well-conditioned gradient. In contrast, STGS applies OneHot coding in the feedforward process and directly outputs a binary association vector. For a discrete output, small perturbations in the training process may not change the output, so the gradient can stay unchanged, resulting in less exploration in learning. This in turn affects the generalization of STGS to a larger number of UEs, as confirmed in our numerical validation section (see Figure. \ref{UE rates}).

\subsection{Complexity Analysis of the Proposed GNN Architecture} \label{complexity}
In a graph with $M$ BSs and $K$ UEs, assume the processing time, or the computation complexity, of each MLP in the GNN is $T$. The value of $T$ is different for training and for inference due to the amount of computation needed (training consists of both forward and backward propagations whereas inference only has forward computation), but can be assume to be fixed in each process. Consider one data sample going through the GNN, the computation times for preprocessing and postprocessing layers are $(M+K+MK)T$ and $2MKT$, respectively. For $L$ update layers, the computation time for updating the node representations at all BS and UE nodes is $L \times 2MKT^2$ while the computation time for all edge representation updates is $L \times MKT$. Thus, the total computation time for node and edge updates is $L(2MKT^2+MKT)$. As such, the total computation time or complexity for processing one data sample through the GNN is $L(2MKT^2+MKT) + (M+K+3MK)T$. As an example, the training time comparison for the GNN with STGS is shown in Table \ref{tab:training time table}.

\setlength{\textfloatsep}{5pt}
\begin{table}[!t]
\caption{STGS GNN Training Time Comparison for One Million Samples}\label{tab:training time table}
\centering
\renewcommand{\arraystretch}{1.0}
\small 
\begin{tabular}{|c||c|}
\hline
M = 2, N = 4, K = 8 & 39 h \\
\hline
M = 4, N = 4, K = 16 & 164 h\\
\hline
M = 6, N = 4, K = 24 & 379 h\\
\hline
\end{tabular}
\end{table}

\vspace{-0.4em}

\section{Simulation Results}
\subsection{Simulation Setup}
In this section, we demonstrate the performance and superiority of the proposed GNN for multi-user multi-cell MIMO beamforming network. We use the carrier frequency of 28GHz and a bandwidth of 1GHz, with the noise power spectral density of $-174$dBm/Hz. We simulated network with 2 BSs and a varying number of UEs uniformly distributed in a region of 200m$\times$200m. Each BS employs 4 antennas arranged in a uniform linear array with transmit power of 30dBm. We adopt the mmWave channel model \cite{capacitypaper}  
\begin{align}
    \mathbf{h}(t)=\frac{1}{\sqrt{I}}\sum_{i=1}^I \alpha_{i}(t)e^{j\phi_{i}(t)}\mathbf{a}_{tx}(\theta^{tx}_{i})
\end{align}
where $\alpha_{i}(t)$ is the path gain on the $i$-th path and $\phi_{i}(t)$ is the uniformly distributed phase. $\mathbf{a}_{tx}(\cdot) \in \mathbb{C}^{N\times 1}$ is the response vector for transmitter antenna array.  

For the GNN, each MLP has 2 hidden layers with 1024 nodes and are trained using the Adam optimizer. Based on our extensive hyperparameter tuning, we set each training epoch to use 400 mini-batches with 5 samples per mini-batch to achieve the best performance. Each sample is created independently where the channels and the locations of BSs and UEs are randomly generated. We assume perfect channel information and use $L=2$ updating layers in GNN. The dimension of each node/edge representation set to 512, we configure the wireless network at 2 BSs and 8 UEs. Then during testing and evaluation, the numbers of UEs are set to be larger to evaluate the GNN scalability performance. We test the average performance of 3000 samples which are again independently generated and have not been seen in training. We choose the mean aggregation. For $\mathbf{d}_\text{GS}$ in Eq. (\ref{d_GS}), the temperature factor $\tau=1$ which was found via extensive testing to be the optimal. For best performance, the GNNs were trained with a warm restart schedule \cite{WR} with an initial restart period of 50 epochs and a multiplier 2, while recycling the learning rate within range [$1\times10^{-8}$, $5 \times 10^{-5}$]. All the experiments are implemented using Pytorch on a NVIDIA A100 GPU.

\subsection{Numerical Results}
\subsubsection{Convergence behaviors of the proposed GNN with different association methods}
Fig. \ref{fig:Convergence} compares the convergence of the proposed GNN using STGS and softmax-based association. Since softmax produces fractional outputs, we apply one-hot encoding by selecting the maximum association value per UE to compute the sum-rate for fair comparison. The GNNs use a warm restart learning rate (restarted every 50 epochs \cite{WR}), causing periodic fluctuations in training and testing performance. This strategy helps the optimizer escape local minima and enhances generalization.
Results show that the GNN with STGS outperforms the softmax variant. This is because STGS produces integer-like outputs during inference while retaining differentiability, guiding the model to learn near-discrete associations. In contrast, softmax yields continuous outputs, leading to suboptimal performance.


\subsubsection{Association factor visualization}
\begin{figure}[!t]
\centering
\includegraphics[width=2.5in]{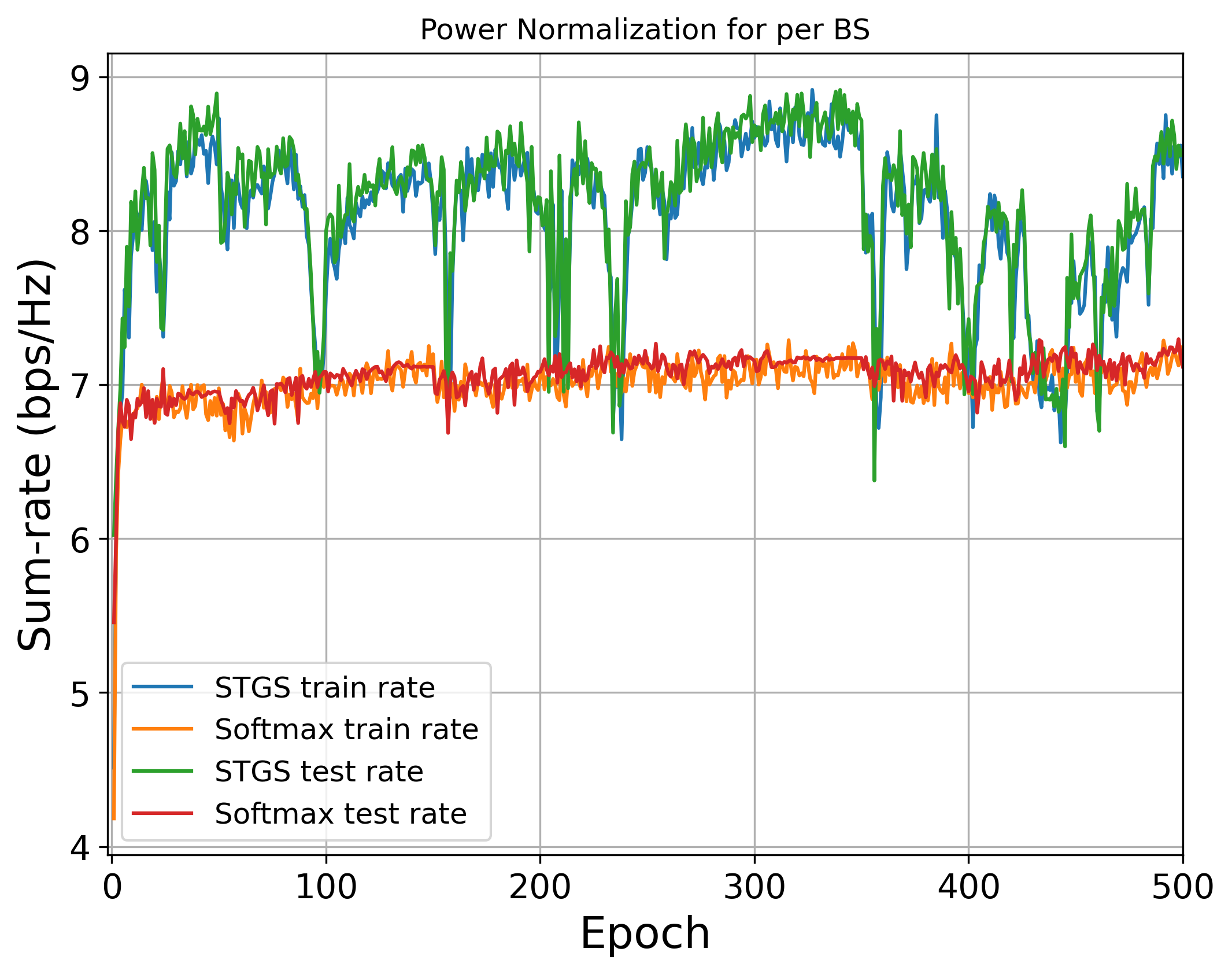}
\caption{Convergence behaviors for GNN models with STGS reparameterization and with $\operatorname*{softmax}$ postprocessing trained with 8 UEs and 2 BSs.}
\label{fig:Convergence}
\end{figure}

\begin{figure}[htbp]
    \centering
    \subfloat[]{%
        \includegraphics[width=0.365\columnwidth]{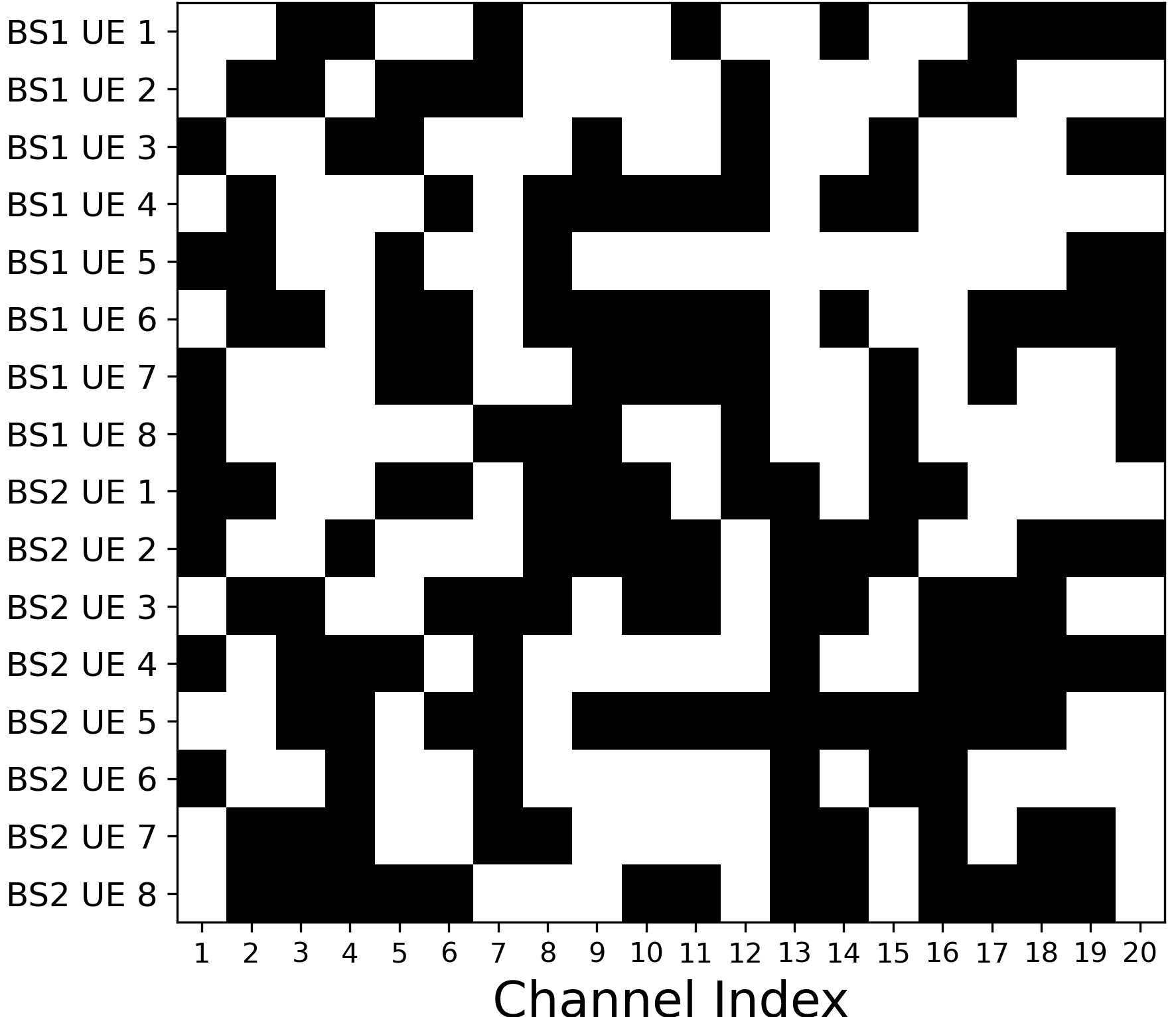}
        \label{fig:STGS_channel}
    }
    \subfloat[]{%
        \includegraphics[width=0.31\columnwidth]{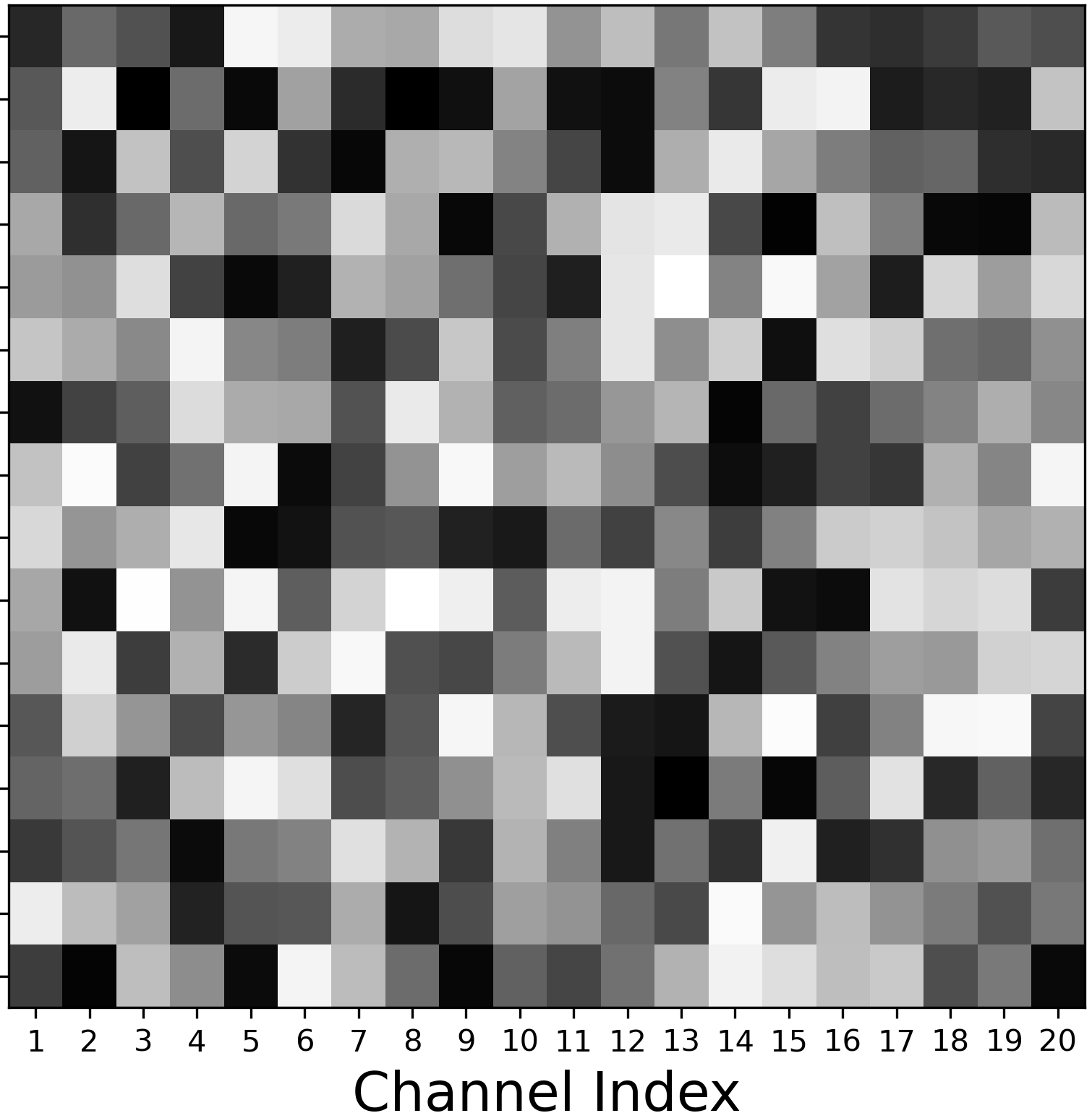}
        \label{fig:GSonly_channel}
    }
    \subfloat[]{%
        \includegraphics[width=0.36\columnwidth]{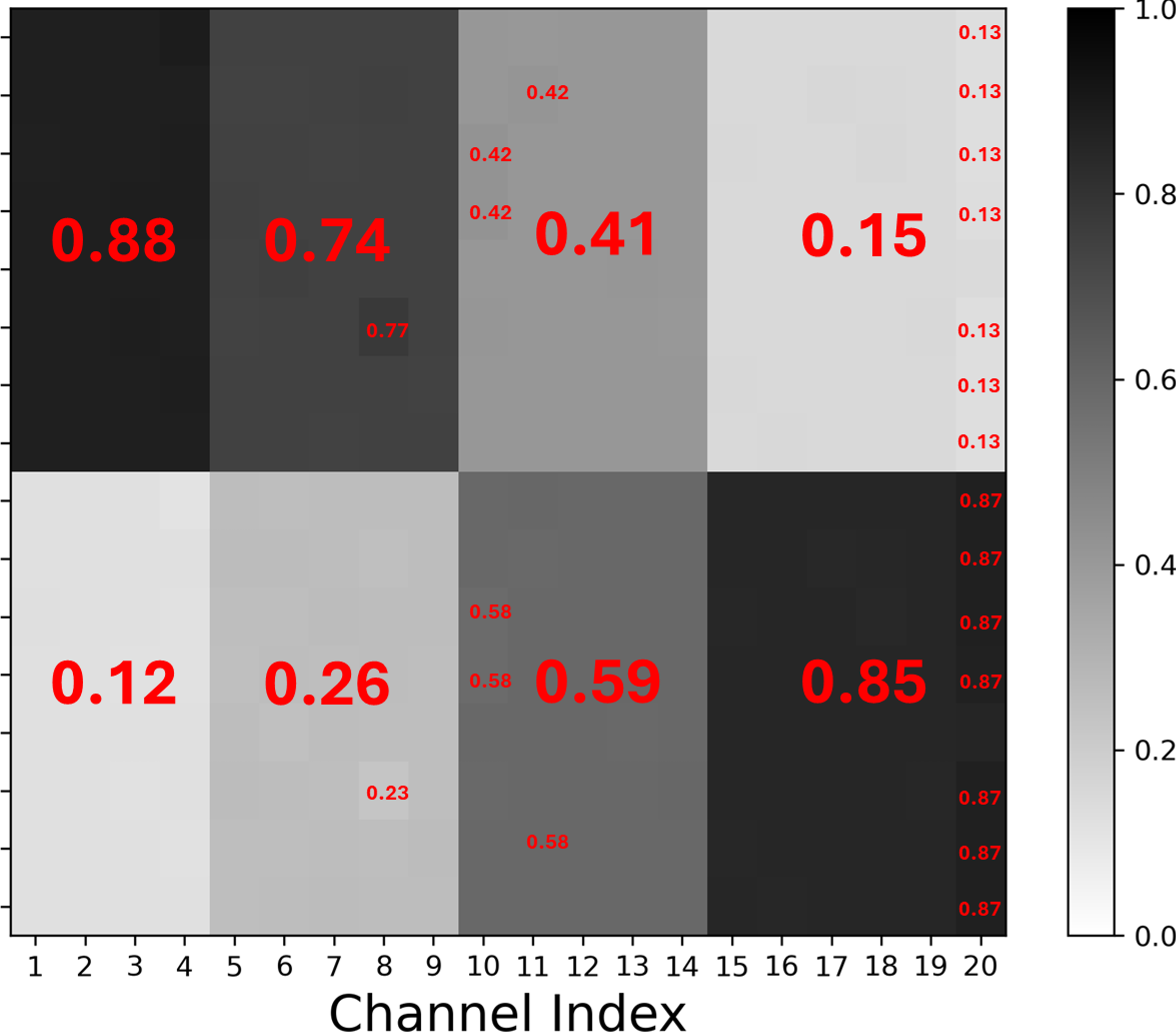}
        \label{fig:softmax_channel}
    } 
    \caption{User association factors from 20 random channels. (a) Integer association factors for the GNN using STGS reparameterization (black = 1, white = 0); (b) Fractional association factors for the GNN using GS reparameterization only; (c) Fractional association factors for the GNN using $\operatorname*{softmax}$ postprocessing (values show the probability of association).}
    \label{fig:combined channel}
\end{figure}

After the GNN is trained, to visualize the association results we randomly select 20 channels and evaluate the association factor outputs for the proposed GNN with STGS and that with $\operatorname*{softmax}$ in Fig. \ref{fig:combined channel}. As shown in Fig. \ref{fig:STGS_channel}, it is clear that the association factors resulting from STGS are always integer. The association factors obtained by using $\operatorname*{softmax}$, shown in Fig. \ref{fig:softmax_channel}, are always non-integer and each UE tends to have similar association decision for all the considered channel samples. This is because $\operatorname*{softmax}$ has no integer restriction during training, and can only produce outputs in the continuous domain. Converting these $\operatorname*{softmax}$ continuous values to integer associations will result in degradation in the network sum-rate compared to what the GNN was trained for, which explains the difference in performance observed in both Fig. \ref{fig:Convergence} and \ref{fig:combined channel}.

\begin{figure}[!t]
\centering
\includegraphics[width=2.5in]{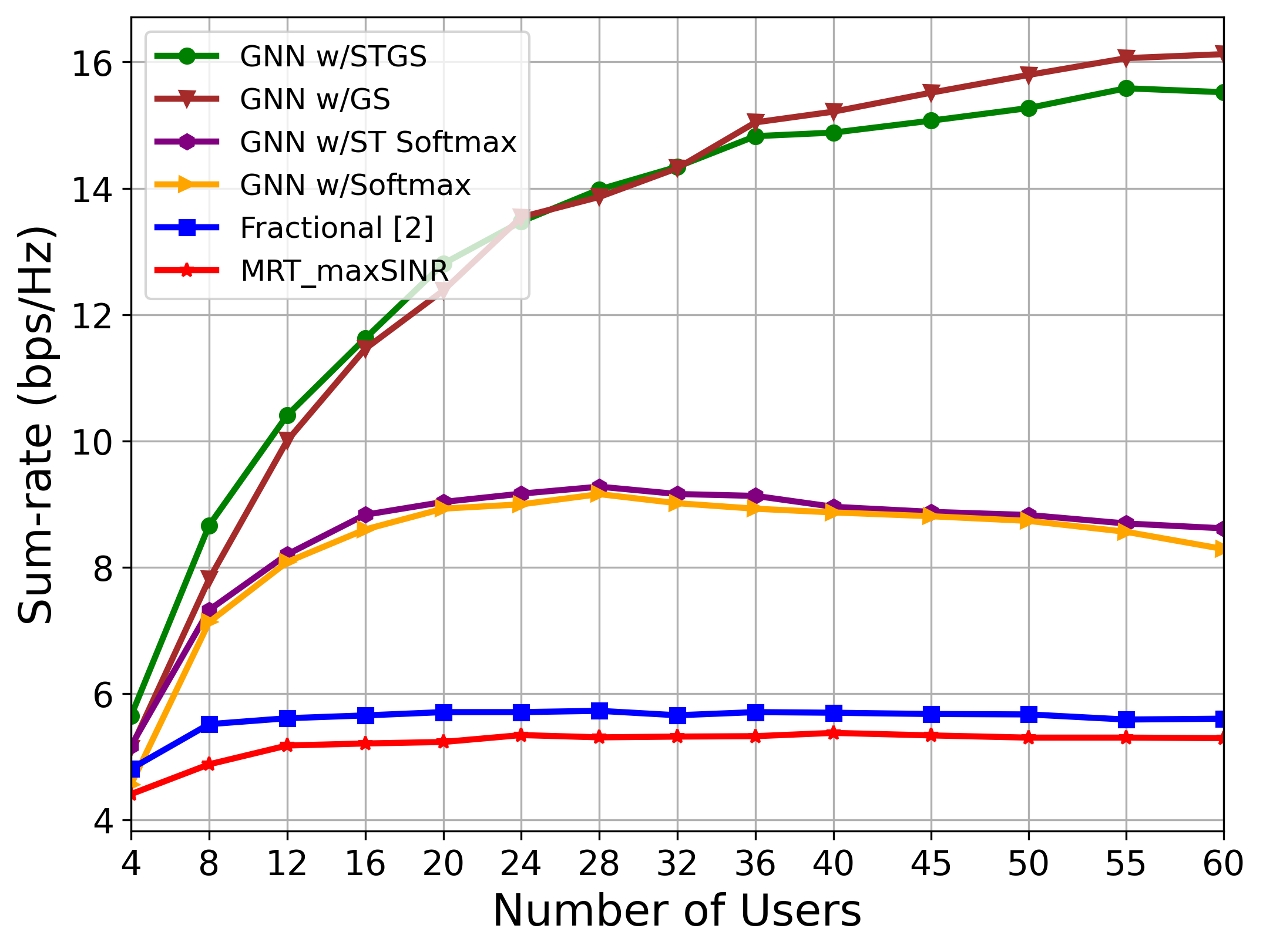}
\caption{Generalization behaviors for GNN models trained with 8 UEs and 2 BSs, then applied to networks with more UEs without retraining.}
\label{UE rates}
\end{figure}

\begin{figure}
    \centering
    \includegraphics[width=2.5in]{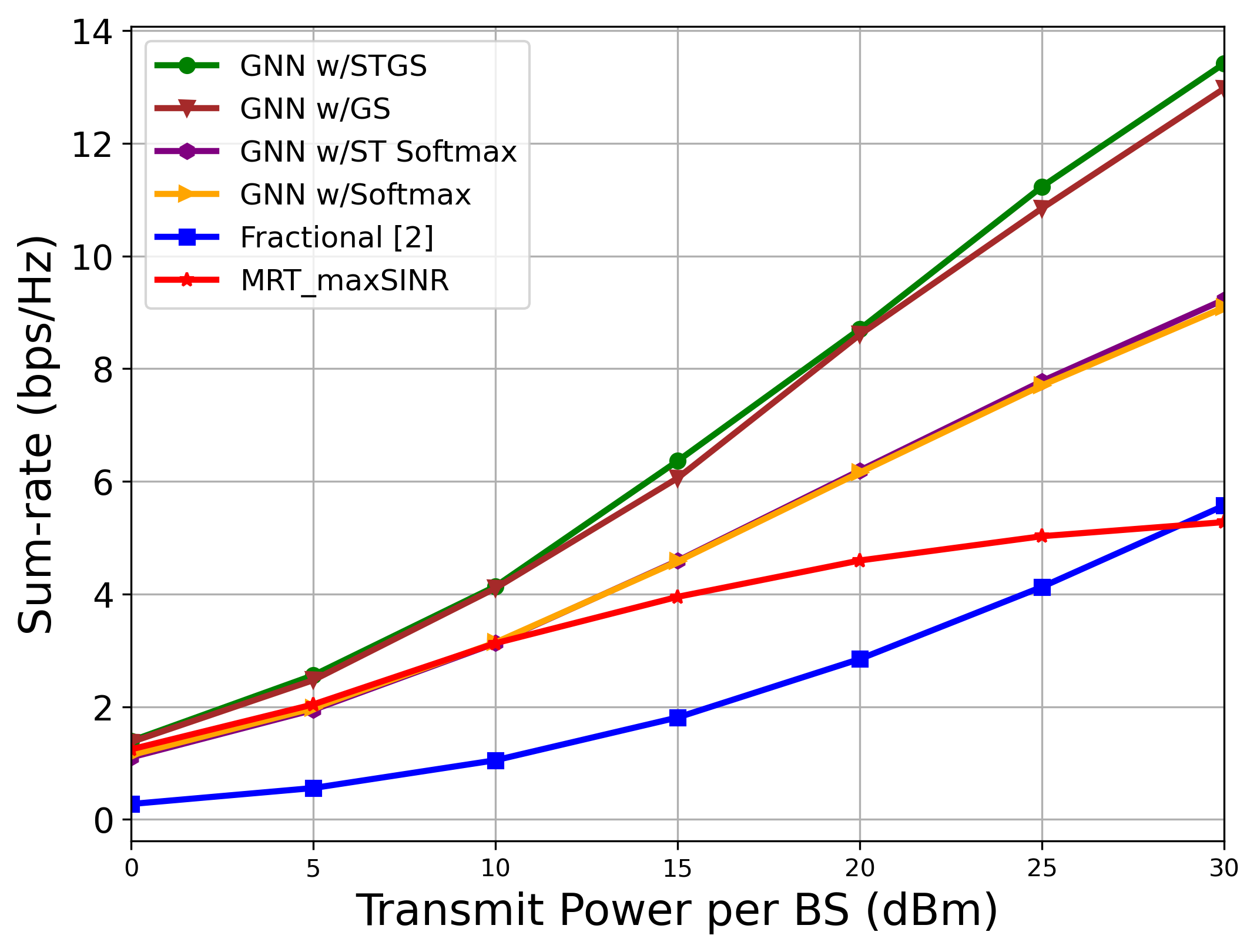}
    \caption{Generalization of GNN models to different power levels. The GNN models were trained with 8 UEs and 2 BSs, each BS with transmit power 30 dBm, and then applied to cases with 24 UEs and various transmit power levels without retraining.}
    \label{Pt rates}
\end{figure}

\begin{figure}
    \centering
    \includegraphics[width=2.5in]{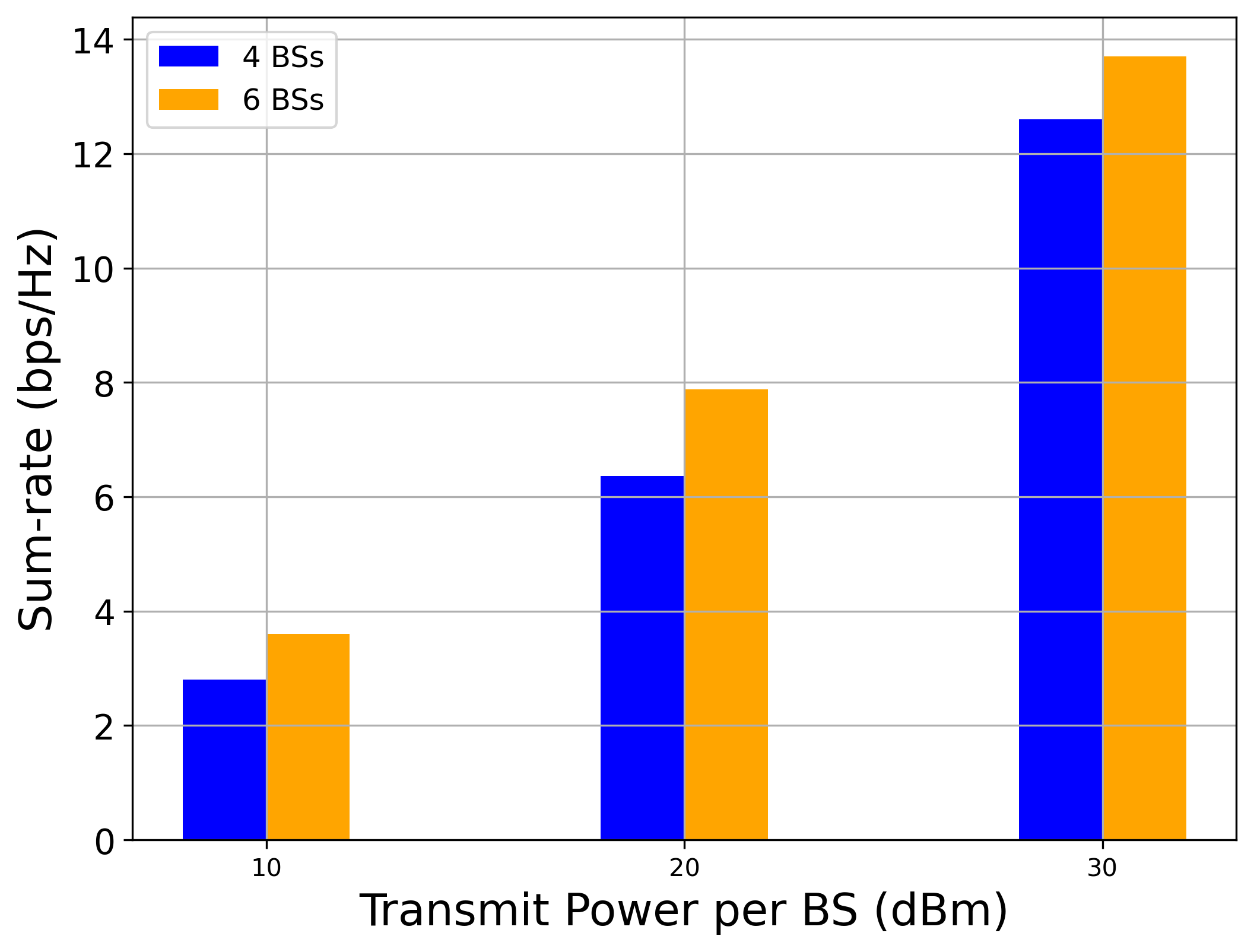}
    \caption{Performance comparison with different BSs and power settings. The GNN models were trained on configurations with 4 BSs and 16 UEs, and with 6 BSs and 24 UEs, respectively, with each BS operating at 30 dBm. They were then generalized to scenarios with 16 UEs and various transmit power levels.}
    \label{BS rates}
\end{figure}

\subsubsection{Generalization abilities}
We further test the generalization ability of the proposed GNN. Here we only consider the generalization on the number of UEs. The GNN can be generalized to a much larger number of UEs, which is practical, as the number of UEs often changes rapidly while the number of BSs remains unchanged.

We compare GNN-based performance with conventional optimization benchmarks. For GNN-based methods, we use the proposed GNN structure with different functions for association decisions. To ensure fairness, non-integer associations are used in the loss for training, while their one-hot versions evaluate the sum-rate. For conventional methods, we consider two baselines: 1) $\text{MRT}\_{\text{maxSINR}}$ where maximum-ratio transmission (MRT) is used for beamforming, and then max-SINR between UE and BSs is used for association; 2) Fractional association solution as in \cite{UAFractional}.  

As shown in Fig. \ref{UE rates}, with the increasing of number of UEs, our proposed GNN with STGS and GS achieve better performance in most cases compared to all the other strategies. STGS is likely to handle user association and resource allocation more effectively due to its ability to directly produce discrete decisions while still being differentiable during training. When there are a large number of UEs (32 UEs), the GS generalization ability becomes slightly better than STGS. Even though GS outputs non-integer association during training, this allows the GNN to potentially learn and explore more, leading to slightly better generalization with more UEs.  

The results can be divided into 3 groups: GS with and without ST, $\operatorname*{softmax}$ with and without ST, and traditional methods. Notably, the performance difference between the variants with and without ST within the GS and softmax groups is minimal. The main benefit of ST is that it produces integer outputs directly. STGS produces slightly better sum rates at training in both groups, but can limit the generation ability to a large number of UEs as observed in Fig. \ref{UE rates}.

Both the STGS and GS methods introduce Gumbel noise based on $\operatorname*{softmax}$, hence allowing the GNN model to explore and can more effectively allocate resource when the network becomes more congested with users than all other schemes. Note that the $\operatorname*{softmax}$ method alone actually degrades performance with an increasing number of UEs. Both traditional $\text{MRT}\_{\text{maxSINR}}$ and Fractional methods do not scale with increasing users as their resource allocation does not effectively capture the interaction among UEs and BSs. This interaction is well captured in the GNN model based on the underlying graph structure and the update mechanism.

Furthermore, we consider the sum-rate versus the transmit power at each BS $P_m$. As shown in Fig. \ref{Pt rates}, the GNN was trained on a network with 8 UEs and 2 BSs at 30 dBm, and tested on scenarios with 24 UEs under various transmit powers. Our proposed GNN models with STGS and GS consistently outperform other methods, even at low power levels. This strong generalization is due to the GNN's ability to capture key network dynamics during training, enabling adaptation to different interference and density conditions without retraining. These results highlight the robustness and practicality of our approach in dynamic wireless environments.

Next, we evaluate performance under different network configurations. Since the number of BSs defines the association matrix dimension, it must remain fixed during inference unless the model is retrained—an increasingly time-consuming process as the number of BS grows. We train two STGS-based GNN models: one with 4 BSs and 16 UEs, and another with 6 BSs and 24 UEs, both at 30 dBm per BS. As shown in Fig. \ref{BS rates}, both models generalize well across various transmit power levels and BS count.

\vspace{-0.6em}
\section{Conclusion}
We proposed a novel GNN architecture to jointly optimize beamforming and integer-based user association in a wireless network. To overcome the challenge that neural networks usually cannot output integer values directly, as these values have zero gradients during backpropagation, we employ Gumbel-Softmax reparameterizations with or without Straight-Through coding to guarantee integer association. Simulation results demonstrate that the proposed reparameterized GNN not only can produce integer association decisions, but also achieves higher sum-rate when generalized to larger networks and various transmit power levels than all other fractional association solutions.

\vspace{-0.7em}
\bibliographystyle{IEEEtran}
\bibliography{ref}

\begin{thebibliography}{10}
\providecommand{\url}[1]{#1}
\csname url@samestyle\endcsname
\providecommand{\newblock}{\relax}
\providecommand{\bibinfo}[2]{#2}
\providecommand{\BIBentrySTDinterwordspacing}{\spaceskip=0pt\relax}
\providecommand{\BIBentryALTinterwordstretchfactor}{4}
\providecommand{\BIBentryALTinterwordspacing}{\spaceskip=\fontdimen2\font plus
\BIBentryALTinterwordstretchfactor\fontdimen3\font minus \fontdimen4\font\relax}
\providecommand{\BIBforeignlanguage}[2]{{%
\expandafter\ifx\csname l@#1\endcsname\relax
\typeout{** WARNING: IEEEtran.bst: No hyphenation pattern has been}%
\typeout{** loaded for the language `#1'. Using the pattern for}%
\typeout{** the default language instead.}%
\else
\language=\csname l@#1\endcsname
\fi
#2}}
\providecommand{\BIBdecl}{\relax}
\BIBdecl

\bibitem{WMMSE}
Q.~Shi, M.~Razaviyayn \emph{et~al.}, ``An iteratively weighted mmse approach to distributed sum-utility maximization for a mimo interfering broadcast channel,'' \emph{IEEE Trans. Signal Process}, vol.~59, no.~9, pp. 4331--4340, 2011.

\bibitem{UAFractional}
Q.~Ye, B.~Rong, Y.~Chen, M.~Al-Shalash, C.~Caramanis, and J.~G. Andrews, ``User association for load balancing in heterogeneous cellular networks,'' \emph{IEEE Transactions on Wireless Communications}, vol.~12, no.~6, pp. 2706--2716, 2013.

\bibitem{2018relational}
P.~W. Battaglia, J.~B. Hamrick \emph{et~al.}, ``Relational inductive biases, deep learning, and graph networks,'' \emph{arXiv preprint arXiv:1806.01261}, 2018.

\bibitem{deng2023gnn_aided}
W.~Deng, Y.~Liu, M.~Li, and M.~Lei, ``Gnn-aided user association and beam selection for mmwave integrated heterogeneous networks,'' \emph{IEEE Wireless Communications Letters}, 2023.

\bibitem{he2022gblinks}
S.~He, S.~Xiong \emph{et~al.}, ``Gblinks: Gnn-based beam selection and link activation for ultra-dense d2d mmwave networks,'' \emph{IEEE Transactions on Communications}, vol.~70, no.~5, pp. 3451--3466, 2022.

\bibitem{fan2022soft}
T.-H. Fan and Y.~Wang, ``Soft actor-critic with integer actions,'' in \emph{2022 American Control Conference (ACC)}.\hskip 1em plus 0.5em minus 0.4em\relax IEEE, 2022, pp. 2611--2616.

\bibitem{jang2017categorical}
E.~Jang, S.~Gu, and B.~Poole, ``Categorical reparametrization with gumble-softmax,'' in \emph{International Conference on Learning Representations (ICLR 2017)}.\hskip 1em plus 0.5em minus 0.4em\relax OpenReview. net, 2017.

\bibitem{revirewer_TCOM}
S.~Zhang, S.~Zhang, F.~Gao, J.~Ma, and O.~A. Dobre, ``Deep learning optimized sparse antenna activation for reconfigurable intelligent surface assisted communication,'' \emph{IEEE Transactions on Communications}, vol.~69, no.~10, pp. 6691--6705, 2021.

\bibitem{RIS}
B.~Lim and M.~Vu, ``Graph neural network based beamforming and ris reflection design in a multi-ris assisted wireless network,'' in \emph{2023 IEEE SSP}, 2023, pp. 120--124.

\bibitem{capacitypaper}
M.~R. Akdeniz \emph{et~al.}, ``Millimeter wave channel modeling and cellular capacity evaluation,'' \emph{IEEE journal on selected areas in communications}, vol.~32, no.~6, pp. 1164--1179, 2014.

\bibitem{WR}
I.~Loshchilov and F.~Hutter, ``Sgdr: Stochastic gradient descent with warm restarts,'' \emph{arXiv preprint arXiv:1608.03983}, 2016.

\end{thebibliography}

\end{document}